\documentclass[aps,pra,twocolumn,showpacs,final,superscriptaddress]{revtex4}
\usepackage{amssymb,amsmath,mathrsfs}
\usepackage{graphicx}

\providecommand{\abs}[1]{{\lvert{#1}\rvert}}
\providecommand{\bra}[1]{{\langle{#1}\rvert}}
\providecommand{\ket}[1]{{\lvert{#1}\rangle}}
\providecommand{\bracket}[2]{{\langle{#1}|{#2}\rangle}}
\newcommand{\proj}[1]{{\ket{#1}\!\bra{#1}}}
\newcommand{\Hilbert}{{\mathscr H}}
\newcommand{\id}{\openone}

\DeclareMathOperator{\supp}{supp}
\DeclareMathOperator{\rank}{rank}
\DeclareMathOperator{\tr}{tr}

\begin{document}
\title{Unambiguous discrimination of mixed quantum states: \newline
       optimal solution and case study}
\author{Matthias Kleinmann}
\email{Matthias.Kleinmann@uibk.ac.at}
\affiliation{Institut f\"ur Theoretische Physik III, 
             Heinrich-Heine-Universit\"at D\"usseldorf,
             40225 D\"usseldorf, Germany}
\affiliation{Institut f\"ur Quantenoptik und Quanteninformation,
             \"Osterreichische Akademie der Wissenschaften,
             6020 Innsbruck, Austria}
\author{Hermann Kampermann}
\author{Dagmar Bru\ss}
\affiliation{Institut f\"ur Theoretische Physik III, 
             Heinrich-Heine-Universit\"at D\"usseldorf,
             40225 D\"usseldorf, Germany}
\begin{abstract}
We present a generic study of unambiguous discrimination between two mixed 
 quantum states.
We derive operational optimality conditions and show that the optimal 
 measurements can be classified according to their rank.
In Hilbert space dimensions smaller or equal to five this leads to the complete 
 optimal solution.
We demonstrate our method with a physical example, namely the unambiguous 
 comparison of $n$ quantum states, and find the optimal success probability.
\end{abstract}
\maketitle

According to the laws of quantum mechanics, two non-orthogonal quantum states 
 cannot be distinguished perfectly.
This fact has far-reaching consequences in quantum information processing, 
 e.g.\ it allows to generate a secret random key in quantum cryptography.
In spite of the fundamental nature of the problem of state discrimination, 
 determining the \emph{optimal} measurement to distinguish two (mixed) quantum 
 states is far from being trivial.

In the literature, two main paths to state discrimination have been taken 
 \cite{Herzog:2004PRA}: firstly, in \emph{minimum error discrimination}, the 
 unavoidable error in distinguishing two states from each other is minimized.
This problem has been completely solved in Ref.~\cite{Helstrom:1976}.
Secondly, in \emph{unambiguous state discrimination} (USD), no error is 
 allowed, but an inconclusive answer may occur.
The optimal USD measurement minimizes the probability of an inconclusive answer 
 \cite{Ivanovic:1987PLA, Dieks:1988PLA, Peres:1988PLA}.
Although USD has found much attention in the recent years, and special examples 
 have been solved, no general solution is known so far for the case of mixed 
 states.
A strategy that is analogous to USD, but applicable also for linearly dependent 
 states are maximum confidence measurements, discussed in 
 Ref.~\cite{Croke:2006PRL}.

The aim of the current contribution is to present the optimal USD measurement 
 for cases which cannot be reduced to the pure state case and thus to known 
 solutions.
This analysis can be applied for the unambiguous discrimination of \emph{any} 
 two density operators acting on a Hilbert space up to dimensions five.
This goes beyond previous results which require a high symmetry or other very 
 special properties of the given states \cite{Bergou:2003PRL, Raynal:2003PRA, 
 Bergou:2005PRA, Herzog:2005PRA, Raynal:2005PRA, Herzog:2007PRA, 
 Raynal:2007PRA}.
We will show the main ideas and steps towards the solution, and explain the 
 technical details elsewhere \cite{Kleinmann:2008AOP}.

The scenario of optimal unambiguous discrimination of two density operators is 
 as follows: two (normalized) density operators $\varrho_1$ and $\varrho_2$, 
 acting on a finite-dimensional Hilbert space $\Hilbert$ occur with \emph{a 
 priori} probability $p_1$ and $p_2$, respectively, where $p_1+p_2=1$.
We will denote the support of a density operator $\varrho$ as the 
 orthocomplement of its kernel, ${(\supp\varrho)}^\perp= \ker\varrho$.
A measurement for USD is described by a positive operator valued measure 
 (POVM), i.e., a family of positive semi-definite operators $\{E_1, E_2, E_?\}$ 
 with $E_1+E_2+E_?=\id$, obeying the constraints for unambiguity, 
 $\tr(E_2\varrho_1)= 0$ and $\tr(E_1\varrho_2)= 0$.
The operator $E_?$ corresponds to the inconclusive outcome while $E_1$ and 
 $E_2$ correspond to the successful detection of $\varrho_1$ and $\varrho_2$, 
 respectively.
The aim is to find a POVM which maximizes the success probability 
 $P_\mathrm{succ}= p_1 \tr(E_1\varrho_1)+ p_2 \tr(E_2\varrho_2)$.
Let us introduce here the useful notation $\gamma_1= p_1\varrho_1$ and 
 $\gamma_2= p_2\varrho_2$.
Thus, the success probability reads $P_\mathrm{succ}= \tr(E_1\gamma_1) + 
 \tr(E_2\gamma_2)$.

What are the relevant structures of the density operators and measurement 
 operators?
The unambiguity condition $\tr (E_2\gamma_1)=0$ means that the support of $E_2$ 
 is a subspace of the kernel of $\gamma_1$.
The second unambiguity condition reads $\supp E_1 \subset \ker \gamma_2$.
Obeying these constraints, one has to maximize the sum of the scalar products 
 $\tr(E_1\gamma_1)$ and $\tr(E_2\gamma_2)$, while keeping $E_?$ positive.
Due to the reduction theorems in Ref.~\cite{Raynal:2003PRA}, the optimization 
 problem reduces to the case of a \emph{strictly skew} pair of (unnormalized) 
 density operators.
The operators $\gamma_1$ and $\gamma_2$ are called strictly skew, when they 
 neither possess any parallel component, i.e., $\supp \gamma_1 \cap \supp 
 \gamma_2= \{0\}$, nor any orthogonal components, i.e., $\supp \gamma_1 \cap 
 \ker \gamma_2=\{0\}$ and $\supp\gamma_2 \cap \ker\gamma_1= \{0\}$.
A simple example for a strictly skew pair of unnormalized density operators is 
 any pair of pure states, $\gamma_1= p\proj{\phi_1}$ and $\gamma_2= 
 (1-p)\proj{\phi_2}$, with $0< \abs{\bracket{\phi_1}{\phi_2}}< 1$ and $0< p< 
 1$.
Both operators of such a strictly skew pair have the same rank, and the sum of 
 both ranks cannot exceed the dimension of the underlying Hilbert space.
---
Below we will show a constructive method to discriminate two skew density 
 operators of rank two.
This solves optimal USD  in all cases where one of the given states has rank 
 two, and hence in articular the case with a Hilbert space of dimension smaller 
 equal five.

In the following we will only consider skew pairs of unnormalized density 
 operators and proper USD measurements.
We call a USD measurement \emph{proper}, if it satisfies $\supp (E_1+E_2) 
 \subset \supp (\gamma_1+\gamma_2)$.
It is sufficient to only consider proper measurements, since the subspace $\ker 
 \gamma_1\cap \ker\gamma_2$ cannot contribute to the success probability 
 \cite{Rudolph:2003PRA}.

In Ref.~\cite{Eldar:2004PRA} Eldar and collaborators showed that the optimality 
 of a USD measurement can be proved via the existence of a certain operator 
 that fulfills a set of conditions.
However, no constructive way to find this operator was provided.
Starting from these conditions we derive the following set of necessary and 
 sufficient requirements for the optimality of a proper USD measurement:
\begin{subequations}\label{e4321}
\begin{gather}
 E_? (\gamma_2-\gamma_1)E_?(\id-E_?)=0, \label{cond1} \\
 \Lambda_1 E_? (\gamma_2-\gamma_1)E_?\Lambda_2= 0, \label{cond2} \\
 \Lambda_1 E_? (\gamma_2-\gamma_1)E_?\Lambda_1\ge 0, \label{cond3} \\
 \Lambda_2 E_? (\gamma_1-\gamma_2)E_?\Lambda_2\ge 0 \label{cond4}.
\end{gather}
\end{subequations}
Here, $\Lambda_1$ is the projector onto $\ker \gamma_2$, and $\Lambda_2$ is the 
 projector onto $\ker \gamma_1$.
The details of the derivation are presented elsewhere \cite{Kleinmann:2008AOP}.
Note that the methods used in order to arrive at Eqns.~\eqref{e4321} cannot be 
 generalized to the discrimination of more than two states.
(For special cases, however, cf.\ Ref.~\cite{Kleinmann:2008XXX}.)

Let us point out two observations from Eqns.~\eqref{e4321}: first, neither 
 $E_1$ nor $E_2$, but only the operator $E_?$ appears in this set of equations.
This is due to the fact, that from $E_?$ it is possible to uniquely reconstruct 
 $E_1$ and $E_2$, as $E_i\gamma_i = \gamma_i-E_?\gamma_i$ holds for $i= 1, 2$.
Second: neither $\gamma_1-\gamma_2 \ge 0$ nor $\gamma_2-\gamma_1 \ge 0$ can 
 hold for a strictly skew pair of operators, and thus it is non-trivial to 
 fulfill Eq.~\eqref{cond3} and Eq.~\eqref{cond4}.
The set of equations \eqref{cond1}--\eqref{cond4} provides an efficient tool in 
 optimal USD\@: one might be able to guess a measurement, e.g.\ from the 
 symmetry of a given USD problem, and can then verify easily whether it is 
 optimal.
Moreover, one can use these equations in a constructive way in order to find 
 the solution for $E_?$, which then uniquely defines an optimal POVM\@.
Below, we will show explicitly how to construct the optimal measurement from 
 Eqns.~\eqref{e4321} for the example of state comparison.

It has been an open question whether the optimal USD measurement is unique.
This is indeed the case.
The structure of the proof is as follows:
As pointed out above, a USD measurement is already defined via $E_?$.
It can be shown \cite{Kleinmann:2008AOP} that for optimal proper measurements 
 the rank of $E_?$ is fixed, namely $\rank E_?= \rank(\gamma_1\gamma_2) +\dim 
 \ker(\gamma_1+\gamma_2)$.
Assuming that there would be two optimal operators $E_?$ and $E_?'$, their 
 convex combination $\frac12(E_?+ E_?')$ would also describe an optimal 
 measurement.
However, for positive semi-definite operators $E_?$ and $E_?'$, the identity 
 $\rank (E_?+ E_?')= \rank E_?= \rank E_?'$ can only hold if $\supp E_?= \supp 
 E_?'$.
When the support of $E_?$ is given, the operator $E_?$ is uniquely determined 
 via Eq.~\eqref{cond1}.
Thus, the optimal proper USD measurement is unique.

\begin{figure}
 \includegraphics[width=50ex]{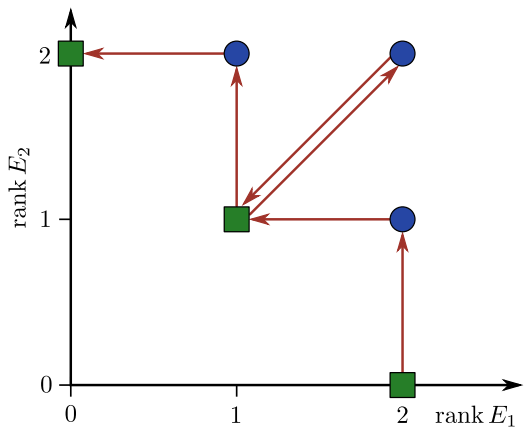}
 \caption{\label{figtypes}
USD measurement types for $r=2$, as allowed by the constraint in 
 Eq.~\eqref{e18835}.
Projective measurements are indicated by squares, non-projective ones by 
 circles.
The arrows illustrate an example for a possible  path between the measurement 
 types, while the probability $p_2$ is varied from $p_2=0$ to $p_2=1$.
The start point is necessarily type $(2,0)$ and the end point type is $(0,2)$.
The types $(0,2)$, $(2,0)$, and $(2,2)$ will only occur once.
Which other types are visited inbetween, and in which order, depends on the 
 concrete example.}
\end{figure}

The uniqueness of the optimal measurement now allows a meaningful 
 characterization of the optimal USD measurement.
We introduce a classification of the different types of optimal USD 
 measurements according to the rank of the measurement operators $E_1$ and 
 $E_2$.
A measurement type is specified by $(\rank E_1, \rank E_2)$.
This classification turns out to be vital for the construction of optimal 
 measurement strategies from Eqns.~\eqref{e4321}.
For given density operators $\varrho_1$ and $\varrho_2$ and a given \emph{a 
 priory} probability $p_1= 1-p_2$, one particular measurement type is optimal, 
 due to the uniqueness of the optimal solution.
While varying $p_1$ some or all of these measurement types may occur, see 
 Fig.~\ref{figtypes} for an illustration.
With $r=\rank \gamma_1= \rank \gamma_2$, one arrives at the constraints $\rank 
 E_1\le r$, $\rank E_2\le r$, and
\begin{equation}\label{e18835}
 r\le \rank E_1+ \rank E_2 \le 2r.
\end{equation}
Eq.~\eqref{e18835} follows from the geometry of unambiguous measurements and 
 the fact that in the optimal case $\rank E_?=\dim \ker \gamma_1\gamma_2$ 
 holds.
The two extremal cases where either the lower or the upper bound in 
 Eq.~\eqref{e18835} is reached correspond to special situations.

The case of the upper bound in Eq.~\eqref{e18835}, where $\rank E_1= r= \rank 
E_2$, is the well-understood \emph{fidelity form measurement}:
Intuition might tell that the success probability should be a function of some 
 distance measure between the two states (this is indeed true for minimum error 
 discrimination, where the smallest achievable error probability is a function 
 of the trace distance between the unnormalized density operators).
Here, for the case with $\rank E_1= r= \rank E_2$ the success probability is 
 the square of the Bures distance, i.e., $P_\mathrm{fid}= 
 1-2\tr\abs{\sqrt{\gamma_1}\sqrt{\gamma_2}}$ \cite{Rudolph:2003PRA, 
 Herzog:2005PRA, Raynal:2005PRA, Kleinmann:2008AOP} (while, in general, 
 $P_\mathrm{fid}$ is an upper bound on the success probability 
 \cite{Rudolph:2003PRA}).
In fact, formally the construction of the fidelity form measurement is always 
 possible \cite{Raynal:2005PRA} and the resulting operator $E_?$ always 
 satisfies all conditions in Eqns.~\eqref{e4321}.
However, this operator in general fails to satisfy the condition $\id- E_?\ge 
 0$.
The measurement types for which $\rank E_1+ \rank E_2< 2 r$ occur due to this 
 very positivity condition.
In a geometric language the optimal measurement is on the border of the allowed 
 (positive) measurements, unless $\rank E_1= r= \rank E_2$.
One can compute two numbers $p_\mathrm{low}$ and $p_\mathrm{up}$ for given 
 $\varrho_1$ and $\varrho_2$, such that the fidelity form measurement is 
 optimal if and only if $p_\mathrm{low}\le p_1 \le p_\mathrm{up}$.

In the case of the lower bound of Eq.~\eqref{e18835}, where $\rank E_1+ \rank 
 E_2= r$, the operators $E_1$, $E_2$, and $E_?$ are projectors, i.e., the 
 optimal measurement is a von-Neumann measurement.
A special situation occurs when $\rank E_1= 0$ and $\rank E_2= r$ or $\rank 
 E_1=r$ and $\rank E_2= 0$.
This is interpreted as follows:
For very small $p_1$ it will turn out to be advantageous to ignore $\varrho_1$ 
 by choosing $E_1=0$.
This case is referred to as \emph{single state detection} of $\varrho_2$, 
 because the state $\varrho_1$ is never detected.
As then $E_?=\id- E_2$, from Eqns.~\eqref{e4321} only Eq.~\eqref{cond3} 
 remains, and this inequality can be written as
\begin{equation}\label{ineq:gamma}
 \gamma_1(\gamma_2-\gamma_1)\gamma_1 \ge 0.
\end{equation}
The success probability for single state detection of $\gamma_2$ is given by 
 $P_\mathrm{succ}=\tr(\Lambda_2\gamma_2)$, where $\Lambda_2$ was defined above 
 as the projector onto $\ker \gamma_1$.
Eq.~\eqref{ineq:gamma} implicitly defines a calculable threshold for $p_1$, 
 below which it is advantageous not to detect $\varrho_1$.
This threshold is always larger than 0, i.e., single state detection is always 
 optimal for a finite regime.
Analogous considerations hold for small $p_2$.

So far our considerations have been independent of $r$.
Let us now consider specific values for $r$.
For $r=1$, i.e., the case of pure states, only the single state detection 
 measurement or the fidelity form measurement may occur.
Hence the problem of unambiguous discrimination of pure states is well 
 understood \cite{Jaeger:1995PLA}.
Furthermore, any USD task where the two density operators can simultaneously be 
 brought in a diagonal form with $2\times 2$-dimensional blocks (the 
 ``block-diagonal'' case), can also be solved by treating the corresponding 
 orthogonal subspaces independently \cite{Raynal:2005PRA, Bergou:2005PRA, 
 Kleinmann:2007JPA}.
For all other cases only solutions for special cases are known 
 \cite{Raynal:2005PRA, Raynal:2007PRA, Herzog:2005PRA, Herzog:2007PRA}.
---
For $r=2$ there are six possible measurement types which are summarized in 
 Table~\ref{table1}.
The optimal measurements for the types $(1,2)$, $(2,1)$, and $(1,1)$ remain to 
 be determined.
For each of these types, the Eqns.~\eqref{e4321} reduce to a polynomial 
 equation \cite{Kleinmann:2008AOP} and hence the analytic solution for the case 
 $r=2$ is completed.

\begin{table}\begin{tabular}{c|c|c|l}
 $\rank E_1$ & $\rank E_2$  &  type & properties\\
 \hline
 0 & 2 & $(0,2)$ & single state detection, projective\\
 1 & 2 & $(1,2)$ & non-projective measurement\\
 2 & 2 & $(2,2)$ & fidelity form measurement, non-proj.\\
 1 & 1 & $(1,1)$ & projective measurement, cf.\ example\\
 2 & 1 & $(2,1)$ & non-projective measurement\\
 2 & 0 & $(2,0)$ & single state detection, projective \\
\end{tabular}
\caption{\label{table1}
Measurement types for the case $r=2$.
For details about the properties see main text.
}
\end{table}

Let us now study the important example of quantum state comparison and 
 demonstrate explicitly how to solve Eqns.~\eqref{e4321} for the case of 
 measurement type $(1,1)$ which occurs for a wide range of parameters.
We consider state comparison of $n$ pure quantum states, where each of the 
 states is taken from the set $\{\ket{\psi_1}, \ket{\psi_2}\}$, with 
 corresponding \emph{a priori} probabilities $\{\eta_1, \eta_2\}$, 
 $\eta_1+\eta_2=1$.
In quantum state comparison \cite{Barnett:2003PLA, Rudolph:2003PRA, 
 Chefles:2004JPA, Herzog:2005PRA, Kleinmann:2005PRA, Kleinmann:2007JPA} one 
 aims at answering the question whether the given $n$ quantum states are equal 
 or not.
Applications of this task in quantum information are e.g.\ quantum 
 fingerprinting \cite{Buhrman:2001PRL} and quantum digital signatures 
 \cite{Gottesman:2001XXX}.
For $n=2$ the optimal unambiguous measurement for quantum state comparison has 
 been given in Ref.~\cite{Herzog:2005PRA, Kleinmann:2005PRA}.
For $n\ge 3$, the corresponding USD task reduces to the unambiguous 
 discrimination of two mixed states of rank 2, i.e., $r=2$.

State comparison of $n$ states is equivalent to the discrimination of (cf.\ 
 Ref.~\cite{Kleinmann:2005PRA})
\begin{align}
 \gamma_\mathrm{e}&= (\eta_1 \proj{\psi_1})^{\otimes n}+
                     (\eta_2 \proj{\psi_2})^{\otimes n},\\
 \gamma_\mathrm{d}&= (\eta_1\proj{\psi_1}+ \eta_2\proj{\psi_2})^{\otimes n}
                     -\gamma_\mathrm{e}.
\end{align}
Due to Theorem~2 in Ref.~\cite{Raynal:2003PRA} it remains to consider the 
 reduced operators $\gamma_\mathrm{e}^r$ and $\gamma_\mathrm{d}^r$, which are 
 given by the projection of $\gamma_\mathrm{e}$ and $\gamma_\mathrm{d}$ onto 
 $(\supp\gamma_\mathrm{e}+ \ker\gamma_\mathrm{d})$, respectively.
It is straightforward to see that for $n\ge 3$ this discrimination task cannot 
 be reduced further and that no block-diagonal structure is present unless 
 $\eta_1= \eta_2= \frac12$.

We next construct a basis of $\supp\gamma_\mathrm{e}$ and of $\ker 
 \gamma_\mathrm{d}$.
A convenient basis of $\supp\gamma_\mathrm{e}$ is given by
\begin{equation}
 \ket{\phi_\pm}\propto \ket{\psi_1}^{\otimes n}\pm
                  \ket{\psi_2}^{\otimes n}.
\end{equation}
We define $c=\bracket{\psi_1}{\psi_2}$ with $0<c<1$.
Using $\ket{\psi_1^\perp}\propto \ket{\psi_2}-c\ket{\psi_1}$ and 
 $\ket{\psi_2^\perp}\propto \ket{\psi_1}-c\ket{\psi_2}$, a basis of 
 $\ker{\gamma_\mathrm{d}}$ can be constructed as
\begin{equation}
 \ket{\omega_\pm}\propto \ket{\psi_1^\perp}^{\otimes n}\pm
                  \ket{\psi_2^\perp}^{\otimes n}.
\end{equation}

Now a Gram-Schmidt orthogonalization of $\{\ket{\phi_+}, \ket{\phi_-}, 
 \ket{\omega_+}, \ket{\omega_-}\}$ yields the orthonormal basis 
 $\{\ket{\phi_+}, \ket{\phi_-}, \ket{\sigma_+}, \ket{\sigma_-}\}$ of 
 $\supp\gamma_\mathrm{e}+ \ker \gamma_\mathrm{d}$.
Then $\{\ket{\sigma_+}, \ket{\sigma_-}\}$ is an orthonormal basis of 
 $\ker\gamma_\mathrm{e}^r\cap \supp(\gamma_\mathrm{e}^r+ \gamma_\mathrm{d}^r)$ 
 while $\{\ket{\omega_+}, \ket{\omega_-}\}$ is an orthonormal basis of 
 $\ker\gamma_\mathrm{d}^r\cap \supp(\gamma_\mathrm{e}^r+ \gamma_\mathrm{d}^r)$.
In fact, they form Jordan bases (cf.\ e.g.\ Ref.~\cite{Stewart:1990, 
 Rudolph:2003PRA}) of these subspaces, i.e., $\bracket{\sigma_\mp}{\omega_\pm}= 
 0$.
The remaining overlaps $\bracket{\sigma_\pm}{\omega_\pm}$ are equal for odd $n$ 
 (\emph{degenerate Jordan angles}).
We now study for general but odd $n\ge 3$ the solution of the conditions in 
 Eqns.~\eqref{e4321} while restricting our considerations to the measurement 
 type $(1,1)$.

The measurements of type $(1,1)$ are von-Neumann measurements, where 
 $E_\mathrm{e}$ and $E_\mathrm{d}$ both have rank 1, i.e., $E_\mathrm{e}= 
 \proj{\chi_\mathrm{e}}$ and $E_\mathrm{d}= \proj{\chi_\mathrm{d}}$.
In particular the vectors $\ket{\chi_\mathrm{e}}$ and $\ket{\chi_\mathrm{d}}$ 
 must be orthogonal and normalized.
We use the parametrization $\ket{\chi_\mathrm{e}}\propto \ket{\omega_+}+ x^* 
 \ket{\omega_-}$ and $\ket{\chi_\mathrm{d}}\propto x\ket{\sigma_+}- 
 \ket{\sigma_-}$ where $x$ is a complex variable \footnote{
This parametrization does not include the case $\ket{\chi_\mathrm{e}}= 
 \ket{\omega_+}$, $\ket{\chi_\mathrm{d}}= \ket{\sigma_-}$.
However, this case is optimal only if $\eta_1= \eta_2= \frac12$.}.

We now evaluate the necessary and sufficient conditions for optimality in 
 Eqns.~\eqref{e4321}:
Eq.~\eqref{cond1} is satisfied for any $x$.
Let us abbreviate $\bra{\omega_a}\gamma_\mathrm{e}\ket{\omega_b}= 
 G_\mathrm{e}^{ab}$ and $\bra{\sigma_a}\gamma_\mathrm{d}\ket{\sigma_b}= 
 G_\mathrm{d}^{ab}$, where $a, b\in \{+, -\}$.
Eq.~\eqref{cond2} now becomes a scalar equation which is only quadratic in $x$; 
 in matrix notation Eq.~\eqref{cond2} reads
\begin{equation}
 (1, x)(G_\mathrm{e}-G_\mathrm{d})(-x,1)^T=0.
\end{equation}
Similarly, the positivity conditions \eqref{cond3} and \eqref{cond4} simplify 
 to scalar inequalities.

\begin{figure}
 \includegraphics[width=55ex]{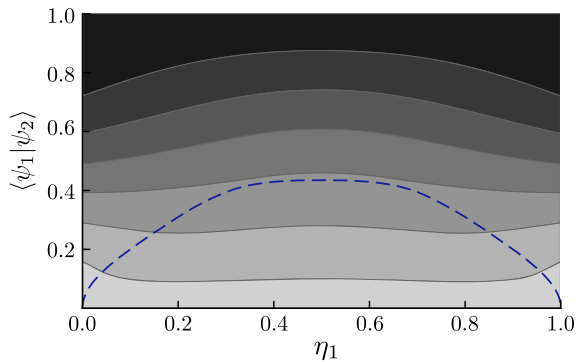}
 \caption{\label{fcont}
 Maximal success probability for comparison of 3 pure quantum states, taken 
  from the set $\{\ket{\psi_1}, \ket{\psi_2}\}$, as a function of the \emph{a 
  priori} probability $\eta_1$ and the overlap $\bracket{\psi_1}{\psi_2}$.
 Darker areas correspond to lower success probability.
 The dashed line indicates the bound from the conditions \eqref{cond3} and 
\eqref{cond4}.
}
\end{figure}
With the help of a computer algebra system we obtain for $n=3$ the optimal 
success probability
\begin{multline}\label{e21347}
 P_\mathrm{succ}^{(1,1)}=\frac14\frac{(1-c^2)^2}{1-c^6}\Big\{
      (c^4+4c^2+1)\,\alpha+\\+(1-c^2)(2+ \sqrt W\,)\Big\},
\end{multline}
 with $W=[(1-c^6)\alpha ^2+4(1-\alpha -\alpha c^4)]
 (1-\alpha)+\alpha^2 c^2$ and $\alpha=4\, \eta_1\eta_2$.
Note that this expression is only valid if in addition the inequalities 
 \eqref{cond3} and \eqref{cond4} hold.
The success probability is illustrated as contour plot in Fig.~\ref{fcont}.
Above the dashed line the optimal measurement is of type $(1,1)$ and the 
 success probability is given by Eq.~\eqref{e21347}.
We find from numerical analysis that the optimal measurement is a fidelity form 
 measurement in the remaining cases.
Note, that for a wide range of the parameters the optimal measurement is a 
 von-Neumann measurement and hence may be implemented physically without the 
 need of an auxiliary system.

In summary, we have presented a strategy to find the optimal measurement for 
 unambiguous discrimination of two mixed quantum states acting on a 
 five-dimensional Hilbert space.
Our method can in principle also be applied to the discrimination of two 
 quantum states in general dimensions.
Our results are also useful in other contexts, e.g.\ quantum state filtering: 
 in Ref.~\cite{Bergou:2003PRL} it has been shown how to optimally distinguish 
 between one pure state from a given set and the remaining ones.
With our method one could filter a subset of states from the whole set.
In connection to quantum algorithms, one could thus distinguish between two 
 sets of Boolean functions, rather than between one function and a set of 
 functions.
The results presented in this paper could also be used to prove optimality for 
 the universal programmable state discriminator suggested in 
 Ref.~\cite{Bergou:2005PRL}.
As the optimal measurement is unique, the optimal device discussed in 
 Ref.~\cite{Bergou:2005PRL} cannot be simplified.
Furthermore, in Ref.~\cite{Raynal:2005PRA} the importance of unambiguous 
 discrimination in the context of quantum key distribution was shown with 
 particular emphasis on the case of states of rank two.
As an outlook, our strategy seems a promising path for the generalization to 
 unambiguous state discrimination of more than two states.

\begin{acknowledgments}
We acknowledge discussions with T.~Meyer, Ph.~Raynal, and R.~Unanyan.
This work was partially supported by the EU Integrated Projects SECOQC, SCALA, 
 OLAQUI, QICS and by the FWF\@.
\end{acknowledgments}

\bibliography{the}
\end{document}